\documentclass[aps,prr,preprint,groupedaddress,longbibliography,tightenlines,superscriptaddress]{revtex4-2}

\usepackage[utf8]{inputenc} 
\usepackage[T1]{fontenc}    
\usepackage{hyperref}       
\usepackage{url}            
\usepackage{booktabs}       
\usepackage{amsfonts}       
\usepackage{nicefrac}       
\usepackage{microtype}      
\usepackage{graphicx}

\usepackage{amssymb} 
\usepackage{amsmath} 
\usepackage{calrsfs} 
\usepackage{amssymb} 
\usepackage[caption=false]{subfig} 
\usepackage{trackchanges}
\addeditor{edit}

\usepackage{accents} 

\def\bfx{\mathbf{x}}

\def\Wtilde{\widetilde{W}}
\def\w2tilde{\widetilde{w}_2}

\def\half{\frac{1}{2}}

\edef\ignore#1{}

\DeclareMathAlphabet{\pazocal}{OMS}{zplm}{m}{n}
\newcommand{\Red}[1]{\textcolor{black}{#1}} 

\usepackage{ntheorem}
\theoremstyle{break} 

\newtheorem{theorem}{Theorem}

\newtheorem{lemma}{Lemma}
\newtheorem{definition}{Definition}

\begin{document}


\title{A Lyapunov theory demonstrating a fundamental limit on the speed of systems consolidation}


\author{Alireza Alemi}
\email[]{alemi@ucdavis.edu}
\affiliation{Center for Neuroscience, and Department of Neurobiology, Physiology, and Behavior, University of California, Davis, Davis, CA 95616, USA}

\author{Emre R. F. Aksay}
\affiliation{Institute for Computational Biomedicine and Department of Physiology and Biophysics, Weill Cornell Medical College, New York, NY 10021, USA}

\author{Mark S. Goldman}
\email[]{msgoldman@ucdavis.edu}
\affiliation{Center for Neuroscience, and Department of Neurobiology, Physiology, and Behavior, University of California, Davis, Davis, CA 95616, USA}
\affiliation{Department of Ophthalmology and Vision Science, University of California, Davis, Davis, CA 95616, USA}



\begin{abstract}
The nervous system reorganizes memories from an early site to a late site, a commonly observed feature of learning and memory systems known as systems consolidation. Previous work has suggested learning rules by which consolidation may occur. Here, we provide conditions under which such rules are guaranteed to lead to stable convergence of learning and consolidation. We use the theory of Lyapunov functions, which enforces stability by requiring learning rules to decrease an energy-like (Lyapunov) function. 
We present the theory in the context of a simple circuit architecture motivated by classic models of \Red{cerebellum-mediated learning and consolidation}.
 Stability is only guaranteed if the learning rate in the late stage is not faster than the learning rate in the early stage. Further, the slower the learning rate at the late stage, the larger the perturbation the system can tolerate with a guarantee of stability. We provide intuition for this result by mapping \Red{a simple example} consolidation model to a damped driven oscillator system and showing that the ratio of early- to late-stage learning rates in the consolidation model can be directly identified with the oscillator’s damping ratio. \Red{We then apply the theory to modeling the tuning by the cerebellum of 
a well-characterized analog short-term memory
system, the oculomotor neural integrator, and find similar stability conditions}. This work suggests the power of the Lyapunov approach to provide constraints on nervous system function.

\end{abstract}


\maketitle



\section{Introduction}

Systems consolidation is the process of transferring learned memories
from an early-stage site to a late-stage site \cite{shadmehr1997neural,mcclelland1995,medina2002parallels}
and has been suggested theoretically to enhance the ability of memory
systems to simultaneously learn new associations while protecting previously learned memories from being overwritten \cite{grossberg1987competitive,mcclelland1995}.
Various forms of memories undergo consolidation in different brain
areas. For example, declarative memories initially learned in the hippocampus
get transferred to the neocortex \cite{mcclelland1995,joo2018hippocampal}.
Motor memories initially located in the cerebellar cortex \cite{krakauer2006consolidation}
or the basal ganglia \cite{andalman2009basal} get transferred out
of the early learning site into direct motor pathways. Furthermore,
strong evidence suggests that fear-based memories initially learned
in the amygdala later get transferred to a different site \cite{medina2002parallels,mcgaugh2004amygdala}.
Understanding how neural signals and learning rules orchestrate a
successful memory transfer requires guiding principles to shed light
on the interactions of brain areas and their plasticity rules. Here
we develop a Lyapunov theory that provides a first-principles account
for the speed of consolidation and the robustness of the consolidation
process.

Neural circuits underlying learning face a fundamental challenge common
to many biological and engineered dynamical systems with adaptively
tunable parameters: the concurrent presence of time-varying inputs,
states, and parameters may cause the dynamics to become unstable,
for example, by growing unboundedly or falling into undesirable oscillatory
patterns. In addition, the nervous system abounds with various forms
of noise and disturbances \cite{faisal2008noise}, which may take
the system into undesirable regimes. Thus, not only should the final
desired solution of learning be stable, but also the overall system
should remain stable throughout the process of learning. 

The theory of adaptive control systems has been successful in providing
essential tools, such as the Lyapunov function formalism, for guaranteeing
the stability of learning systems \cite{slotine1991}. The concept
of a Lyapunov function has been used for quantifying the stability
of adaptive recurrent neural networks \cite{sanner95,alemi18,gilra17}
as well as for building attractor neural networks to model
long-term memory \Red{\cite{hopfield1982neural,amit1989modeling,hahnloserSeung2000Nature}}. Here,
we apply the Lyapunov formalism to the problem of guaranteeing stable 
systems consolidation. 
Systems consolidation, like many learning processes, contains feedback loops between the training signals and neural dynamics that drive learning, and the weight changes that drive system dynamics. Such feedback can make learning prone to instability. This may be further exacerbated by the fact that, in biological systems, many synapses do not have direct access to the ground-truth performance error and must therefore learn from indirect error signals.

We place our theory in the context of a simple circuit architecture \Red{for} systems consolidation \Red{consisting of an early-stage and a late-stage learner. Critically,} the late stage, unlike the early stage, lacks direct access to an error-correcting teaching signal (Fig.~\ref{Fig1}).
In this architecture, the early-stage activity trains the weight(s)
of the late \Red{stage}, such that the memory is eventually transferred from
the early to the late \Red{stage}. This architecture and set of learning
rules correspond to classic theories of learning and systems consolidation
in the cerebellum. Learning at the early\Red{-stage} site \Red{of plasticity} corresponds to the classic
Marr-Albus-Ito theory of cerebellar learning \cite{miles1981plasticity,kassardjian2005site}
in which an error signal conveyed by climbing fibers trains the cerebellar
output conveyed by Purkinje cells. These Purkinje cell firing rates
then serve as a secondary teaching signal for the late\Red{-stage} site \Red{of plasticity} located
\Red{in the cerebellar output nuclei}. 
Applying the formalism of Lyapunov stability theory, we find that, to guarantee stability, plasticity
 at the late-stage site must not be updated faster than
plasticity at the early-stage site. \Red{For a given speed of plasticity at the early-stage site,} the slower the tuning in the late
stage, the more robust the learning process is against noise in the
primary teaching signal. \Red{We then show that the theory can be applied to the learning and tuning of a simple recurrent short-term memory network, the oculomotor neural integrator, that is responsible for stably maintaining the position of the eyes in their orbits  \cite{seung1996brain}.}

\section{Results}

\subsection{Toy model for systems consolidation}

We first consider a simple toy model of systems consolidation motivated
by the basic circuitry thought to be involved in systems consolidation
of cerebellum-mediated learning. The task is to track a given time-varying
input command $r_{\text{in}}(t)$ and to generate a desired output
$r_{o}^{*}(t)=w^{*}r_{\text{in}}(t)$ where the desired input-to-output
gain is denoted as $w^{*}$. The model, shown in Fig.~\ref{Fig1},
has an architecture with two pathways: a direct pathway with gain
$w_{2}$ and an indirect pathway with gain $w_{1}$ that provides
a learned, online correction to the output of the direct pathway.
The output of the model can be written as 
\begin{align}
r_{o}(t)=(w_{1}+w_{2})r_{\text{in}}(t)\label{rO}
\end{align}
where $r_{o}(t)$ is the output. 
 The goal of learning is two-fold: (1) to reduce the gain error $\Wtilde=w_{1}+w_{2}-w^{*}$
to zero, and (2) to consolidate the weight changes into the late-stage
weight such that asymptotically $w_{1}=0$ and $w_{2}=w^{*}$. The
indirect pathway receives the information about the error signal and
constitutes the early site of plasticity. The direct pathway constitutes
the late site of plasticity and needs to be tuned based on the information
received from the early stage, i.e., $r_{1}=w_{1}r_{\text{in}}$.

The learning rule in the early stage is a supervised delta-like correlational
rule proportional to the product of the input and the teaching signal
\cite{widrow1990,arnold1997oculomotor}. 
\begin{align}
\dot{w}_{1}=-\eta_{1}r_{\text{in}}(t)(e(t)+\xi(t))\label{w1lr}
\end{align}
where the teaching error signal $e(t)$ is the tracking error $e(t)=r_{o}(t)-r_{o}^{*}(t)=(w_{1}+w_{2}-w^{*})r_{\text{in}}(t)$,
$\eta_{1}$ is the learning rate of the early-stage tuner, and $\xi(t)$
is a perturbation to the error signal. 
\Red{ We assume 
 that the maximal absolute value of the perturbation at any given time point is signal-dependent \cite{harris1998signal,jones2002sources}, i.e., the strength of the perturbation depends on the
magnitude of error signal $e(t)$, and is given by $ |\xi(t)| \leq\mu |e(t)|$  where $\mu>0$ is a constant parameter of the model that defines the regime of possible perturbations. A bounded constant noise component could be added, as illustrated in Fig.~\ref{Fig1}C,D,
 but is smaller for large error signals where the biggest issues are expected to arise (such as the unbounded growth in Fig.~\ref{Fig1}F); thus, it is omitted for simplicity in the analytical calculations. Alternatively, such a constant noise term could be offset by adding a decay term in Eq.~\ref{w1lr} \cite{bhasin2024synaptic}. }  
The learning rule in the late stage is a heterosynaptic correlational
rule between the online corrective signal $r_{1}(t)$ provided by
the early stage, and the direct input to the late stage, $r_{\text{in}}(t)$
\cite{clopath2014cerebellar}: 
\begin{align}
\dot{w}_{2}=\eta_{2}r_{\text{in}}(t)r_{1}(t)\label{w2lr}
\end{align}
where $\eta_{2}$ is the learning rate.

\begin{figure}[htb]
\includegraphics[scale=0.86]{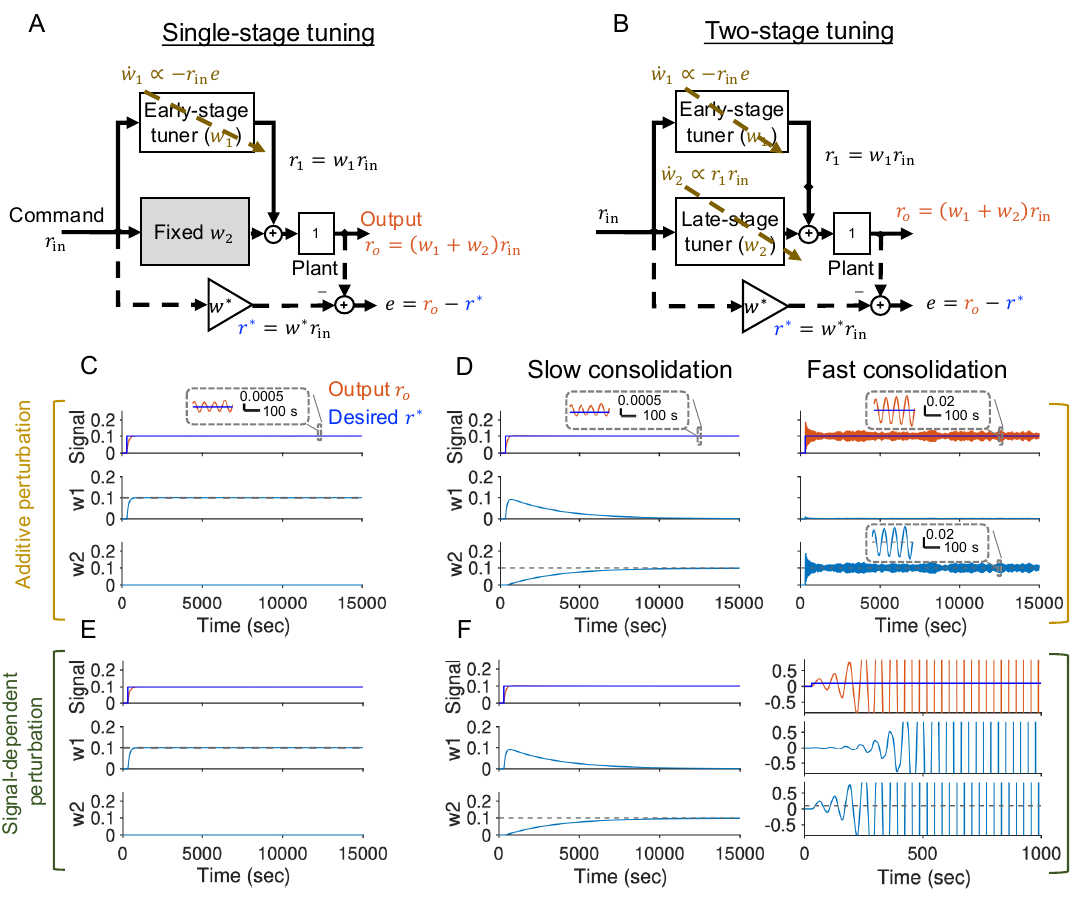}\label{Fig1}
\caption{\label{Fig1}A toy tracking model demonstrating instability in systems
consolidation.~(A)~Top: Single-stage model. The early-stage
parameter, $w_{1}$, is directly tuned by the error signal $e(t)$,
whereas the late-stage parameter, $w_{2}$, is fixed.  ~(B)~Two-stage
tuning model. The early stage is as in the single-stage model. The late-stage weight $w_{2}$ is tuned using the output of the early stage as a secondary teaching signal.
\Red{~(C)~ Simulation of single-stage model} showing that the model successfully converges to the desired output $r^*$ and the desired, tuned weight $w_1$ (dashed line) \Red{in the case of additive perturbation.} 
\Red{~(D)~ In the two-stage model with additive perturbation,} when the consolidation process is slow enough (left panel), the model dynamics successfully converge
and tune the weight $w_2$ to its desired value (dashed line). However, if the consolidation process becomes too fast (right panels), the system can show instability. 
\Red{~(E,F)~ Similar to (C,D) but for a signal-dependent perturbation for which fast consolidation leads to unbounded grouth. Note the abbreviated time scale of the right panel in (F), chosen for better visibility of the unbounded growth.}
See Appendix~A for simulation details.
}
\end{figure}

We seek a general framework to characterize the conditions for stability
and convergence of this two-stage learning system. More specifically,
we seek criteria for tuning the system to avoid undesired behaviors
of the system, such as oscillation or unbounded growth. By stability,
we intuitively mean that, as time goes to infinity, the system is
well-behaved; for example, it should reach its goal and stay close
to it. The mathematical definition of Lyapunov stability and its various
refined notions are provided in Appendix~B.

\subsection{Stability of the single-stage \Red{toy} model}

To illustrate the Lyapunov stability approach and motivate the problem
of instability in systems consolidation, we
first consider a one-stage model (Fig.~\ref{Fig1}A) in which
learning occurs only at
the early stage of the circuit (i.e., we set $\frac{dw_2}{dt} = 0$).
In this case, so long as the magnitude of the perturbation $\xi(t)$ does not exceed the magnitude of the error signal $e(t)$, the learning is guaranteed to stably converge. This is illustrated for the task of learning to track a step-like input in Fig.~\ref{Fig1}A\Red{,C,E} and proven for the more general case in the following paragraphs. 

The basic idea behind Lyapunov's direct approach to stability is based
on constructing a scalar function and showing that all trajectories
of the dynamics of the system decrease this scalar function, guaranteeing
that the system safely and stably reaches the fixed point of the dynamics.
This scalar function, known as a Lyapunov function, can be considered
a generalization of the concept of energy in classical mechanics. Finding one such function
is enough to prove the stability of the system. To provide intuition
about the Lyapunov function formalism, we first apply it to the simple
single-stage tuning model given by Eq.~\ref{w1lr}.

To prove the uniform global stability of the model, we use the Lyapunov
theorem \ref{theorem:Lyapunov} for non-autonomous systems, and to
prove asymptotic stability, we use the Lyapunov-like Lemma~\ref{lemma:lyapunovlike}
(Appendix B). To prove the uniform global stability, we need to find
an appropriate Lyapunov function candidate $L$ such that 1) $L$
is positive definite, 2) the time derivative of $L$ is $\dot{L}\leq0$,
3) $L$ is decrescent, and 4) $L$ at $t=0$ is radially unbounded.
We propose the following scalar function of the error in
the gain $\Wtilde$: 
\begin{equation}
L_{1}=\half\Wtilde^{2},
\end{equation}
where $\Wtilde=w_{1}+w_{2}-w^{*}$. $L_{1}$ is positive definite
and radially unbounded by inspection. To show that $\dot{L}_{1}\le0$,
we write the learning dynamics as $\dot{w}_{1}=-\eta_{1}\Wtilde r_{\text{in}}^{2}-\eta_{1}r_{\text{in}}\xi$.
Using the learning rule to compute the time derivative yields $\dot{L}_{1}=-\eta_{1}\Wtilde^{2}r_{\text{in}}^{2}-\eta_{1}\Wtilde r_{\text{in}}\xi$.
With $|\xi|\leq\mu|e|$, one obtains: 
\begin{equation}
\begin{split}
\dot{L}_1 & \leq-\eta_{1}\Wtilde^{2}r_{\text{in}}^{2}+\eta_{1}|\Wtilde r_{\text{in}}||\xi|\\
 & \leq-\eta_{1}\Wtilde^{2}r_{\text{in}}^{2}+\eta_{1}|\Wtilde r_{\text{in}}|\mu|\Wtilde r_{\text{in}}|\\
 & =-\eta_{1}r_{\text{in}}^{2}(1-\mu)\Wtilde^{2}.
\end{split}
\label{noisyLdot1Stage}
\end{equation}
$\eta_{1}>0$ by definition, and as long as $\mu\leq1$, we have $\dot{L}_{1}\leq0$.
If $r_{\text{in}}=0$, we are in a trivial case where $\dot{L}_{1}=0$
and $\dot{w}_{1}=0$ and the system is not changing at all. We assume
$r_{\text{in}}\neq0$ throughout most of the learning period. Therefore,
the equilibrium is globally stable. This stability guarantees that
$w_{1}$ and $\Wtilde$ are bounded. Since $L_{1}$ does not explicitly
depend on time and is positive definite, it is decrescent; therefore,
the equilibrium is uniformly globally stable. To find conditions for
guaranteeing asymptotic stability, i.e., as $t\rightarrow\infty$,
$w_{1}\rightarrow w^{*}$, from Lemma~\ref{lemma:lyapunovlike} what
is left to show is that $\dot{L}_{1}$ is uniformly continuous
in time. A practical way of showing uniformity is to show that $\ddot{L}_{1}$
is bounded. 
Given that the hyperparameters $\eta_{1},\eta_{2}$ and the gain error
$\Wtilde$ are bounded, and $r_{\text{in}}$ and $\xi$ are assumed
to be smooth functions of time with bounded derivatives, $\ddot{L}_{1}$
is bounded.
  $\blacksquare$


\subsection{Lyapunov function theory for the two-stage \Red{toy} consolidation model}
\label{2stagetoymodel}

We next consider systems consolidation in the two-stage model \Red{(Fig.~\ref{Fig1}B)}. \Red{When} the consolidation process is sufficiently slow,
the two-stage learning model successfully converges to solving the
tracking task (Fig.~\ref{Fig1}\Red{D,F,} left). However, when the
consolidation process is too fast, the system can exhibit instability
in which a small perturbation can cause large perturbations in the
output (Fig.~\ref{Fig1}\Red{D,F,} right). Below, we analytically
find the conditions for guaranteeing stability (Section~\ref{Theory}), provide intuition
for the source of possible instability in regions without stability guarantee by solving
a special case of the system (Section~~\ref{Intuition}), and demonstrate with simulations
a case in which loss of stability leads to unbounded growth of activity
(Section~~\ref{blowup}).

\subsubsection{Theory}
\label{Theory}

\begin{figure}
\includegraphics[scale=0.9 ]{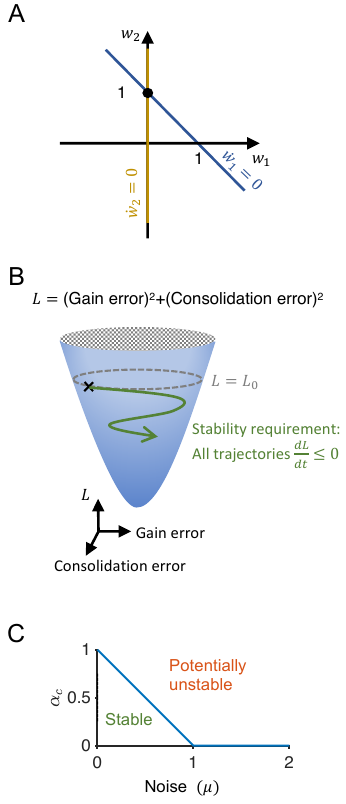}\caption{\label{Fig2} Lyapunov function theory for stability of the 
two-stage \Red{toy} model. ~(A)~ In the limit that the perturbation goes to
zero, $\mu\rightarrow0$, the closed-loop learning dynamics has a
single fixed point and two nullclines (shown for $w^{*}=1$). ~(B)~The
Lyapunov function candidate $L$ has two terms: the squared gain
error and the squared consolidation error. The most important property in
order to have stable convergence in the Lyapunov sense is that the
dynamics of the learning rules should avoid going uphill on the Lyapunov
function surface. ~(C)~When the ratio of learning rates $\alpha=\eta_{2}/\eta_{1}$
is less than a critical value $\alpha_{c}=1-\mu$, the learning is
guaranteed to be stable. As the maximum perturbation amplitude reaches
$|e|$, i.e., $\mu=1$, the region of guaranteed stability vanishes.}
\end{figure}

To investigate the stability and convergence of the two-stage model,
we need to find an appropriate Lyapunov function candidate. 
The two learning rules can be rewritten as $\dot{w}_{1}=-\eta_{1}\Wtilde r_{\text{in}}^{2}-\eta_{1}r_{\text{in}}\xi$
and $\dot{w}_{2}=\eta_{2}w_{1}r_{\text{in}}^{2}$. We choose the following Lyapunov
function candidate for the two-stage model: 
\begin{equation}
L=\half(\Wtilde^{2}+\widetilde{w}_{\text{2}}^{2}),
\end{equation}
where $\widetilde{w}_{\text{2}}=w_{2}-w^{*}$. 
We refer to the first term in the above as the (squared) gain error and the second term as the (squared) consolidation error.
The nullclines and the fixed points of the learning dynamics are shown
in the weight space in Fig.~\ref{Fig2}A for $w^{*}=1$ in the limit
that the amplitude of the perturbation goes to zero, i.e., $\mu\rightarrow0$.
The first term encourages the gain error to go towards zero in a stable
manner and stay close to zero, which is the goal of the learning rule
for $w_{1}$. 
The second term aligns with the goal of consolidating the learned
memories into $w_{2}$ and is achieved when the desired gain $w^{*}$
is only due to $w_{2}$ (Fig.~\ref{Fig2}B). 

We now turn to proving uniform global stability and asymptotic stability
of the two-stage model. Using $w_{1}=\Wtilde-\widetilde{w}_{\text{2}}$,
the time derivative of $L$ can be written as $\dot{L}=\dot{\Wtilde}\Wtilde+\dot{\widetilde{w}}_{2}\widetilde{w}_{2}=-\eta_{1}r_{\text{in}}^{2}((1-\alpha)\Wtilde^{2}+\alpha\widetilde{w}_{\text{2}}^{2})-\eta_{1}\Wtilde r_{\text{in}}\xi$,
where $\alpha=\eta_{2}/\eta_{1}$. With $|\xi|\leq\mu|e|$ one obtains:
\begin{equation}
\begin{split}\dot{L} & =-\eta_{1}r_{\text{in}}^{2}((1-\alpha)\Wtilde^{2}+\alpha\w2tilde^{2})-\eta_{1}\Wtilde r_{\text{in}}\xi\\
 & \leq-\eta_{1}r_{\text{in}}^{2}((1-\alpha)\Wtilde^{2}+\alpha\w2tilde^{2})+\eta_{1}|\Wtilde r_{\text{in}}|\mu|\Wtilde r_{\text{in}}|\\
 & =-\eta_{1}r_{\text{in}}^{2}((1-\alpha-\mu)\Wtilde^{2}+\alpha\w2tilde^{2}).
\end{split}
\label{noisyLdot}
\end{equation}
The main requirement for Lyapunov stability is to show $\dot{L}\leq0$,
which is achieved when $\alpha\leq1-\mu$. As in the single-stage
model, we assume $r_{\text{in}}\neq0$ throughout most of the learning
period. 
For the rest of the proof, we need to verify the other conditions
of Theorem~\ref{theorem:Lyapunov}. $L$ is bounded from below, $\min(L)=0$,
and $L$ does not explicitly depend on time. Hence, $L$ is positive
definite and decrescent. 
We therefore conclude that the equilibrium is uniformly stable.
 Since $L$ is the sum of two quadratic terms, it
is radially unbounded by inspection, which guarantees that $\Wtilde$,
$\w2tilde$, $w_{2}$, $w_{1}$ are globally bounded. To guarantee
asymptotic stability, what is left to show is that $\dot{L}$ is
uniformly continuous in time by showing $\ddot{L}$ is bounded. Given that the hyperparameters $\eta_{1},\eta_{2},\alpha$, and the
variables $\Wtilde$, $\w2tilde$, $w_{1}$ are bounded, and $r_{\text{in}}$
and $\xi$ are assumed to be smooth bounded functions of time with
bounded derivatives, $\ddot{L}$ is bounded.$\blacksquare$

The key result of the above is that, when the late stage is tuned
at a rate not faster than the early stage rate, i.e., $\alpha=\eta_{2}/\eta_{1}\leq\alpha_{c}=1-\mu$,
the system provably remains globally stable and is guaranteed to successfully converge (Fig.~\ref{Fig2}C). Intuitively, in the extreme case where the learning rate of $w_{1}$
is much lower than that of $w_{2}$, it is easy to see why the system
may become unstable: $w_{1}$ moves infinitesimally slowly towards
the goal, but $w_{2}$ gets rapidly updated with a secondary teaching
signal that is not in the direction of the gradient of the error.
This leads to an alteration of the error signal feeding back onto
the early ($w_{1}$) site of learning, potentially causing the learning
process to become unstable. To combat this potential source of stability,
the learning rate at $w_{2}$ should be slower than that of $w_{1}$
to filter out noise and prevent run-away amplification.

When $\alpha>\alpha_{c}$, the analysis only indicates that the system may become prone to instability, but does not itself say whether the system will become unstable. Such instability can potentially arise if a perturbation brings the system into regions where the derivative of $L$ becomes positive, which for this system occurs when $|\widetilde{w}_{\text{2}}|<\sqrt{\frac{ \alpha-\Red{\alpha_c}}{\alpha}}|\Wtilde|$, showing that the size of this region increases with $\alpha$.

To check whether we can improve the stability conditions (i.e., find a higher value of $\alpha_c$) by considering a different relative weighting of the gain and consolidation error terms of $L$, we consider the Lyapunov function $L_{\text{b}}=\half(\Wtilde^{2}+b\widetilde{w}_{\text{2}}^{2})$, where $b>0$. 
For simplicity, we work in the regime $\mu\rightarrow0$, for which $\alpha_{c}=1$.  Calculating the time derivative $\dot{L}_{\text{b}}=-\eta_{1}r_{\text{in}}^{2}({(1-\alpha)}{\Wtilde}^{2}+\alpha(1-b)\Wtilde\widetilde{w}_{\text{2}}+\alpha b{\tilde{w}}_{2}^{2})$, we note that $\dot{L}_{\text{b}}$ is again guaranteed to be less than or equal to zero in the whole weight space as long as $\alpha \leq 1$, but not for $\alpha > 1$ (in particular, this is easily seen when $\tilde{w}_{2}=0$). Thus, the same fundamental criterion for guaranteeing stability emerges even for different weightings of the two error terms of the Lyapunov function $L$.

\subsubsection{Intuition for instability}
\label{Intuition}

\label{SecIntuition}
\begin{figure}
\includegraphics[scale=0.83 ]{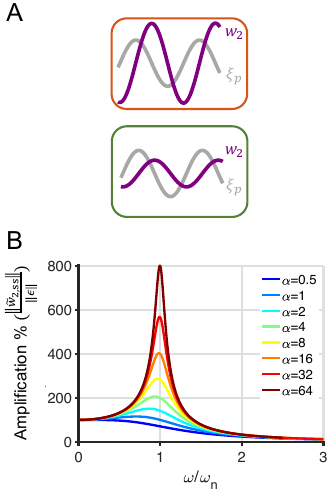}\caption{\label{Fig3} Amplification of perturbation in the region without
\Red{stability guarantee in the two-stage toy model.}
~~(A)~ The \Red{steady}-state of $w_{2}$ exhibits an amplification of  an infinitesimal sinusoidal perturbation probe $\xi_{\text{p}}=\epsilon\sin(\omega_{n}t)$ \Red{when $\alpha > 1$},
 shown \Red{for $\omega = \omega_n$} in the limit that $\mu\rightarrow0$. 
Top, $\alpha=3$ (red box); bottom, $\alpha=0.33$ (green box). ~~(B)~
The steady-state percent amplification of the sinusoidal probe perturbation as a function of the ratio of
the normalized frequency (normalized by the undamped natural frequency
$\omega_{n}$) of the probe $\xi_{p}$ in the limit that
$\mu\rightarrow0$. 
}
\end{figure}

In the $\alpha>\alpha_{c}$ regime, our Lyapunov stability analysis only shows that stability is not guaranteed and thus only indicates the potential for instability. Therefore, 
it is instructive to investigate this regime more closely. Consider
the case where a sinusoidal probe perturbation $\xi_{p}=\epsilon\sin(\omega t)$
with an infinitesimal amplitude $\epsilon$ is present in Eq.~\ref{w1lr}.
 We examine its effect on the system in the regime that $\mu\rightarrow0$
while, for simplicity, we set $r_{\text{in}}=1$. To gain intuition
about this amplification, we solve the system in the presence of the
probe. By eliminating $w_{1}$ in the two learning rules, we obtain
\begin{align}
\ddot{\widetilde{w}}_2 + \eta_1  \dot{\widetilde{w}}_2 + \eta_1 \eta_2 \widetilde{w}_2 = -\eta_1\eta_2\epsilon \sin(\omega t).
\label{seondorderW2}
\end{align}
This second-order differential equation is equivalent to forced mass-spring-damper
dynamics $m\ddot{x}+c\dot{x}+kx=F$, where $x$ is the object displacement,
$m=1$ is the mass, $c=\eta_{1}$ is the damping coefficient, $k=\eta_{1}\eta_{2}$
is the spring constant, and $F=-\eta_{1}\eta_{2}\epsilon\sin(\omega t)$
is the external force. When the damping ratio $\zeta=\frac{c}{2\sqrt{km}}=\frac{1}{2\sqrt{\alpha}}<1$,
 we are in the underdamped regime. 
 The steady-state response, which is dominated by $\widetilde{w}_{2,\text{ss}}(t)$ since $w_1$ approaches zero in the steady state, then has the following form:
\begin{align}
\widetilde{w}_{2,\text{ss}}(t)=-\frac{\epsilon\sin(\omega t+\phi)}{\sqrt{(1-\frac{\omega^{2}}{\omega_{n}^{2}})^{2}+(2\frac{\omega}{\omega_{n}}\zeta)^{2}}},
\end{align}
where $\phi=\arctan{\frac{2\zeta(\frac{\omega}{\omega_{n}})}{1-(\frac{\omega}{\omega_{n}})^{2}}}$
and $\omega_{n}=\sqrt{\eta_{1}\eta_{2}}$ is the undamped natural
frequency.
As $\alpha$ increases, the damping ratio $\zeta$ decreases,
and the amplitude of the $\widetilde{w}_{2,\text{ss}}(t)$ resonance increases.
\Red{At the natural frequency, the amplitude of $\widetilde{w}_{2,\text{ss}}(t)$ equals $\sqrt{\alpha}\epsilon$, showing that a small error perturbation $\xi_{p}$ leads to an amplified output when $\alpha > 1$ (Fig.~\ref{Fig3}A). }
 The mathematical correspondence between the ratio of learning rates $\alpha$ in systems consolidation and the (inverse square of the) damping ratio in a physical oscillator, and the resultant resonant amplification, is the intuition behind the potential instability in the
presence of perturbations. This resonant behavior is shown for a sweep of relative
frequencies for a larger range of $\alpha$ values in Fig.~\ref{Fig3}B.

\label{blowup}

\Red{For the additive perturbation of Fig.~\ref{Fig1}D, }
the resonance caused amplification but this amplification was kept finite by the damping. 
In the presence of signal-dependent perturbation, the natural
frequency of the unperturbed system can interact with the frequency
of vibration of the perturbation. When this interaction gives rise
to a frequency close to the natural frequency of the unperturbed system,
especially when the amplitude of the perturbation is sufficiently
large, there can be increasing amplification of the system in each
period, leading to unbounded growth. This resonance phenomenon, often
referred to as parametric resonance \cite{Verhulst2009}, can happen
when two oscillators get coupled in such a way that one causes oscillations
in the parameters of the other oscillator, and does not necessarily
need an external force to exhibit instability.

\begin{figure}
 \includegraphics[scale=.69]{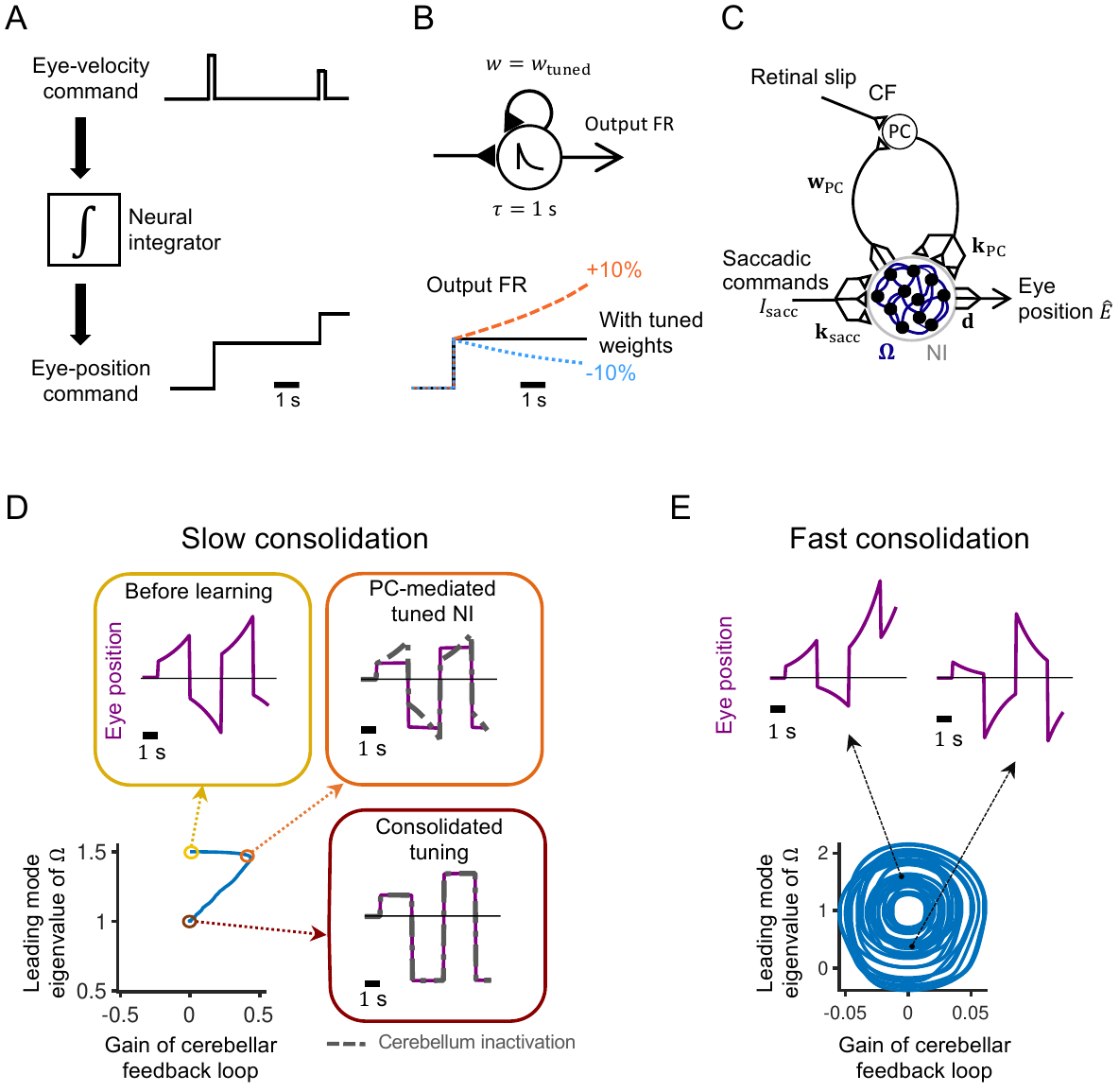}%
\caption{{\label{Fig4} \Red{Cerebellar tuning and subsequent consolidation of the time constant of the oculomotor neural integrator.
~~(A)~The computation of a neural integrator.
~~(B)~The fine-tuning problem of neural integrators. Increasing or decreasing the strength of recurrent feedback by a small amount causes exponential growth or decay, respectively, of integrator activity 
(The shown neuronal time constant $\tau=1$ s is taken from [46], but the results apply more broadly).
 FR: firing rate.
~~(C)~Two-stage model for tuning the time constant of the neural integrator (NI). PC: cerebellar Purkinje cell. CF: climbing fiber carrying the retinal slip error signal.
~~(D)~Tuning the network from an initially unstable integrator condition (yellow), for a model with a fast learning rate in the early site of plasticity (cerebellum) and a slow learning rate in the late site (integrator). In the early stage of learning (orange), behavior is tuned primarily through the plasticity of the early site weights $\mathbf{w}_\text{PC}$, so that the integrator function depends on the cerebellum (dashed grey line: effect of cerebellar inactivation). The learned memory then gets consolidated into the recurrent connectivity $\mathbf{\Omega}$ of the neural integrator circuit, making the function independent of the cerebellum (dark magenta). Lower left panel shows strength of feedback in the cerebellar-NI feedback loop (x-axis), and within the NI recurrent network (y-axis, as characterized by 
the largest eigenvalue of $\mathbf{\Omega}$).
~~(E)~Tuning the model from the same initial condition as in (D)  but with the learning rates of the early and late sites  interchanged,
  so the early site has the slower learning rate.
 This can lead to instability in the system, as demonstrated by this example of oscillatory behavior in the weight space, resulting in alternating unstable (left) and leaky (right) eye positions, with a period ranging from 200 to 280 seconds.}
}
}
\end{figure}

\Red{\subsection{Lyapunov function theory for tuning the recurrent dynamics of the oculomotor neural integrator}}
\Red{We next apply the theory to analyze the two-stage tuning of the dynamics of a recurrently connected neural circuit that is, furthermore, recurrently connected to the cerebellum. We do this within the context of a well-characterized short-term memory circuit of the vertebrate brainstem, the oculomotor neural integrator. This circuit mathematically integrates (in the sense of calculus) motor command signals conveying the desired velocity of the eyes into motor command signals conveying the desired position of the eyes \Red{(Fig.~\ref{Fig4}A)}. In the absence of eye velocity command inputs, network activity is maintained persistently at an analog value \cite{robinson1989integrating,aksay2000anatomy}.
 Dynamically, such persistent activity corresponds to a line of fixed points, or \Red{a} line attractor, that is thought to be created by positive feedback within the network offsetting the inherent tendency of neuronal activity to decay back to a baseline firing level \cite{seung1996brain}. Models of line attractors require precise tuning of their recurrent weights \cite{seung1996brain,seung2000} for this offset to occur (Fig.~\ref{Fig4}B). Within the oculomotor system, 
 this precise tuning might, like other visuomotor control tasks, be driven by visual error-feedback signals conveyed to the cerebellum \cite{major2004a,raymond18}.
However, to our knowledge, no prior work has shown how this tuning can be learned initially by the cerebellum and then consolidated into the oculomotor neural integrator.}

\Red{The cerebellum in this model is recurrently connected to the neural integrator so that its dynamics are tuned via a cerebello-integrator loop  (Fig.~\ref{Fig4}C). Fast plasticity instructed by retinal slip signals conveyed by climbing fibers (CF) occurs in the synapses onto a single lumped cerebellar Purkinje cell (PC) population. On a slower time scale, Purkinje cell activity then tunes the leading eigenmode of the neural integrator network dynamics. We model the neural integrator as a recurrently connected linear network of $N$ neurons \cite{seung1996brain} that is reciprocally connected to the Purkinje cell as follows:}
\Red{\begin{align}
\tau \dot{r}_i &= - r_i + \sum_{j=1}^{N} \Omega_{ij} r_j - k_{\text{PC},i} r_\text{PC} + k_{\text{sacc},i} I_\text{sacc}\label{eqNI1tau}\\
r_\text{PC} &= \sum_{j=1}^{N}  w_{\text{PC},j} \, r_j\label{eqNI2tau}\\
\hat{E} &= \sum_{j=1}^{N}  d_j \, r_j ,\label{eqNI3tau}
\end{align}
where $\tau$ is the time constant of an integrator neuron, $r_i$
 is the firing rate of neuron $i$,  $\Omega_{ij}$ denotes the connection weight from neuron $j$ to neuron $i$, $k_{\text{PC},i}$ is the strength of the input from the cerebellum to neuron $i$, $r_\text{PC}$ is the Purkinje cell firing rate,
  \Red{$k_{\text{sacc},i}$ is the strength of the saccadic input to neuron $i$,}
  $I_\text{sacc}$ is the velocity-encoding eye movement (saccadic) commands, $w_{\text{PC},j}$ is the strength of the plastic synapse onto the Purkinje cell, $\hat{E}$ is the internal representation of the eye position, and $d_j$ is the $j^{th}$ component of the eye position decoder, which is assumed to be fixed during learning.
}

 \Red{We non-dimensionalize time and rates by measuring them in units of the time constant $\tau$ as follows: $ (t/\tau) \rightarrow t$, $ r_i \tau \rightarrow r_i$, and $ I_\text{sacc} \tau \rightarrow I_\text{sacc} $. The resulting dimensionless equations, in vector-matrix notation, are:}

\Red{
 \begin{align}
 \dot{\mathbf{r}} & = - \mathbf{r} +  \mathbf{\Omega} \mathbf{r} - \mathbf{k}_\text{PC} \, \mathbf{w_\text{PC}}^\top \mathbf{r} + \mathbf{k}_\text{sacc} I_\text{sacc}\label{eqNI1}\\
\hat{E} &= \mathbf{d}^\top \mathbf{r} \label{eqNI2}
\end{align}
where $\mathbf{k}_\text{PC}, \mathbf{k}_\text{sacc},  \mathbf{r}, \mathbf{w_\text{PC}}$, and $\mathbf{d}$ are column vectors with $N$ components,  $\mathbf{\Omega}$ is an $N \times N$ matrix with zero diagonal elements, $^\top$ denotes the matrix transpose operation, and $r_\text{PC} = \mathbf{w_\text{PC}}^\top \mathbf{r}$. The desired behavior is perfect integration so that the eye position during eye fixations (in the absence of eye velocity commands)  does not change, i.e., $\dot{E} = I_\text{sacc}$ or $E = \int I_\text{sacc} dt$.
 The retinal slip errors can be written as 
 \begin{align}
\dot{e} & = \dot{\hat{E}} - \dot{E}\nonumber\\ 
	     & = -\mathbf{d}^\top (\mathbf{k}_\text{PC} \mathbf{w}_\text{PC}^\top + \mathbb{I}_N - \mathbf{\Omega} ) \mathbf{r} + (\mathbf{d}^\top \mathbf{k}_\text{sacc}-1)I_\text{sacc},
            \label{RSNI}
\end{align}
where $\mathbb{I}_N$ is the $N\times N$ identity matrix. 
\Red{To achieve the desired behavior, $\dot{e}$ must be zero. From the second term above, this yields $ \mathbf{d}^\top \mathbf{k}_\text{sacc} =1 $}.
 From the first term, denoting the term in parentheses as 
$\widetilde{\mathbf{W}} \equiv  \mathbf{k}_\text{PC} \mathbf{w}_\text{PC}^\top + \mathbb{I}_N - \mathbf{\Omega}$, 
 this yields $\mathbf{d}^\top \widetilde{\mathbf{W}}=0$. 
 Further, denoting a desired solution of $\mathbf{\Omega}$ after consolidation (when $\mathbf {w}_\text{PC} = 0$) as $\mathbf{\Omega}^*$, yields $\mathbf{d}^\top \mathbf{\Omega}^*=\mathbf{d}^\top$, which gives the expected condition that, after consolidation, the neural integrator weight matrix should have a unity eigenvalue. 
 If we define the deviation of $\mathbf{\Omega}$ from its desired solution as ${\widetilde{\mathbf{\Omega}} \equiv \mathbf{\Omega}-\mathbf{\Omega}^*}$,
  then the component of $\widetilde{\mathbf{W}}$
  along the readout vector $\mathbf{d}$ can be decomposed as ${\mathbf{d}^\top \widetilde{\mathbf{W} }= \mathbf{d}^\top (\mathbf{k}_\text{PC} \mathbf{w}_\text{PC}^\top - \widetilde{\mathbf{\Omega}} )}$,
  i.e., as a sum of terms due to the deviation of the neural integrator weight matrix from its desired post-consolidation value $\mathbf{d}^\top \widetilde{\mathbf{\Omega}}$  and due to the (non-zero before consolidation) feedback loop through the cerebellar Purkinje cells.}

\Red{
 The correlational learning rules in this non-dimensionalized notation are:
\begin{align}
\dot{\mathbf{w}}_\text{PC} &=  \eta_1 (\dot{e} + \xi(t)) \mathbf{r}\label{eqNI3}\\
\dot{\mathbf{\Omega}}      &=  -\eta_2 (\mathbf{k}_\text{PC} r_\text{PC}) \mathbf{r}^\top, \label{eqNI4}
\end{align}}
\Red{where $\eta_1$ is the non-dimensionalized learning rate for the early site in the cerebellum, $\eta_2$ is the corresponding learning rate for the late site in the neural integrator, and $\xi(t)$ is a perturbation to the error signal with maximal value bounded by a fraction of the error signal, $ |\xi| \leq\mu | \dot{e}|$.
We consider a Lyapunov function candidate for the two-stage tuning of the neural integrator consisting of the sum of squares of the components of $\mathbf{d}^\top \widetilde{\mathbf{\Omega}}$ and $\mathbf{d}^\top \widetilde{\mathbf{W}}$:
\begin{align}
L &=  \frac{1}{2}(\mathbf{d}^\top \widetilde{\mathbf{W}}) (\mathbf{d}^\top \widetilde{\mathbf{W}})^\top
+ \frac{1}{2}(\mathbf{d}^\top \widetilde{\mathbf{\Omega}}) (\mathbf{d}^\top \widetilde{\mathbf{\Omega}})^\top.
\end{align}}
\Red{As in the Lyapunov function for the two-stage model in the previous section, the first term in $L$ encourages the retinal slip error to vanish, and the second term aligns with the goal of consolidating the learned memories 
from the early stage to the late stage synapses.
}

\Red{We now prove uniform global stability and asymptotic stability
of the two-stage neural integrator tuning model. The time derivative of $L$ is:
\begin{equation}
\begin{split}\dot{L} &=  (\mathbf{d}^\top \dot{\widetilde{\mathbf{W}}}) (\mathbf{d}^\top \widetilde{\mathbf{W}})^\top
+ (\mathbf{d}^\top \dot{\widetilde{\mathbf{\Omega}}}) (\mathbf{d}^\top \widetilde{\mathbf{\Omega}})^\top \\
&=   \mathbf{d}^\top(\mathbf{k}_\text{PC} \mathbf{\dot{w}}_\text{PC}^\top - \dot{\mathbf{\Omega}} ) (\mathbf{d}^\top \widetilde{\mathbf{W}})^\top
+ (\mathbf{d}^\top \dot{\mathbf{\Omega}}) (\mathbf{d}^\top \widetilde{\mathbf{\Omega}})^\top\\
 &=  -\eta_1  (c\,(1-\frac{\alpha}{c})\dot{e}^2 + c\,\xi  \dot{e} + \alpha ( \mathbf{d}^\top \widetilde{\mathbf{\Omega}} \mathbf{r} )^2 ), \\
\end{split}
\label{LdotNI}
\end{equation}
where $\alpha=\eta_2/\eta_1$ and $c=\mathbf{d}^\top \mathbf{k}_\text{PC}$. Similar to the previous case, we can use the defined noise regime to prove $L$ is non-decreasing:
\begin{equation}
\begin{split}
\dot{L} & = -\eta_1 (c\,(1-\frac{\alpha}{c}) \dot{e}^2   + \alpha ( \mathbf{d}^\top \widetilde{\mathbf{\Omega}} \mathbf{r} )^2 ) - \eta_1 c\, \xi  \dot{e}\\
	& \leq -\eta_1 (c\,(1-\frac{\alpha}{c}) \dot{e}^2   + \alpha ( \mathbf{d}^\top \widetilde{\mathbf{\Omega}} \mathbf{r} )^2 ) + \eta_1  c\, \mu | \dot{e}| | \dot{e}|\\
	& = -\eta_1  (c\,(1-\frac{\alpha}{c}-\mu) \dot{e}^2   + \alpha (  \mathbf{d}^\top \widetilde{\mathbf{\Omega}} \mathbf{r} )^2 ).
\end{split}
\label{LdotNIxi}
\end{equation}
When $\alpha\leq (1-\mu) c$, for $c>0$, $\dot{L}\leq0$.
 \Red{If we consider $\mathbf{d}$ and $\mathbf{k}_\text{PC}$ to be normalized to unit length, then $c$ is the cosine of the angle between $\mathbf{d}$ and $\mathbf{k}_\text{PC}$. If $c=1$, i.e., $\mathbf{d}$ and $\mathbf{k}_\text{PC}$ are parallel, then we recover the same condition as in the toy model of Section \ref{2stagetoymodel}. As the angle between $\mathbf{d}$ and $\mathbf{k}_\text{PC}$ increases, $c$ gets smaller, which causes the condition for stability to become tighter.}
We verify the other conditions of Theorem ~\ref{theorem:Lyapunov} as follows:
 $L$ is bounded from below, $\min(L)=0$,
and $L$ does not explicitly depend on time. Hence, $L$ is positive
definite and decrescent.
\Red{Thus,} the equilibrium is uniformly stable.
 Since $L$ is the sum of two quadratic terms, it
is radially unbounded by inspection, which guarantees that $\widetilde{\mathbf{W}}$, $\mathbf{\Omega}$, and $\mathbf{w}_\text{PC}$ are globally bounded. To guarantee
asymptotic stability, what is left to show is that $\dot{L}$ is
uniformly continuous in time by showing $\ddot{L}$ is bounded. Given that the hyperparameters $\eta_{1},\eta_{2},\alpha$, $\tau$,  the vectors $\mathbf{d}$, $\mathbf{k}_\text{PC}$, and the
variables $\widetilde{\mathbf{W}}$, $\mathbf{\Omega}$, $\mathbf{w}_\text{PC}$ are bounded, and $\mathbf{r}$
and $\xi$ are assumed to be smooth bounded functions of time with
bounded derivatives, $\ddot{L}$ is bounded.$\blacksquare$} 

\Red{
Simulations confirm that two-stage tuning of the neural integrator in the presence of signal-dependent noise is stable when learning in the early site is faster than learning in the late site (Fig.~\ref{Fig4}D, Methods). Starting from an initial unstable condition, the simulations successfully converged, and 
 the learned memory was consolidated into the neural integrator. In the early phases of learning, the circuit function is dependent on the cerebellum (Fig.~\ref{Fig4}D, dashed lines). Following consolidation, the neural integrator function becomes independent of the cerebellum. Thus, experiments that inactivate the cerebellum at different times during training of the oculomotor neural integrator should provide testable predictions of this two-stage model. Notably, if the early, cerebellar site of plasticity is slow and the late site of plasticity in the neural integrator is fast, this can potentially cause instability according to the theory. This potential instability is illustrated in Fig.~\ref{Fig4}E, for which the learning rates at the early and late sites have been interchanged relative to that of Fig.~\ref{Fig4}D.}

 
~
\section{Discussion}

We have provided a framework for studying the stability of systems
consolidation and applied it to simple circuit architecture\Red{s} characterized
by an early learning area that is directly trained by performance
errors, which \Red{then} trains a late learning area that provides the final
site of memory storage. Using a Lyapunov function theory that enforces
the stability of the learning and consolidation process, we obtained
a fundamental result on the speed of learning: the late stage must
not be tuned faster than the early stage, and when the teaching signal
is corrupted by perturbation, the late stage should be tuned more
slowly. We mapped the consolidation process \Red{of a simple example circuit} to the dynamics of a driven damped oscillator, providing the intuition that increasing the ratio of late- to early-stage learning rates $\alpha$ is like decreasing the oscillator damping, leading to potential resonant instability. \Red{We then applied the theory to a biologically relevant case of tuning a neural integrator circuit.}

Previous work on memory consolidation has focused primarily on a fundamental
robustness-speed tradeoff in learning, known
as the stability-plasticity dilemma, \Red{in which} having fast plasticity leads to `instability' in the sense that new memories overwrite old ones. \Red{This previous work has shown that the tradeoff can be lessened in multi-stage models \cite{roxin2013efficient,benna2016computational}. 
\Red{Here, within a two-stage model}, we show a complementary, dynamical form of instability that occurs for the systems consolidation of graded memories,}
\Red{ where having an excessively fast consolidation speed}
can lead to amplification of a perturbation or even exponential, unbounded growth
of activity. 

Our toy two-stage model \Red{in Fig.~\ref{Fig1}} maps onto the architecture of classic models
of motor learning mediated by the \Red{cerebellum} \cite{miles1981plasticity,kassardjian2005site}.
In such models, early learning is thought to occur through plasticity
of the weight $w_{1}$ between presynaptic parallel fiber inputs and
postsynaptic Purkinje cells. This plasticity is thought to be driven
by correlations between the activity of parallel fiber inputs
\Red{and} behavioral error signals that are conveyed by separate, climbing
fiber inputs to the Purkinje cells. The learning process is particularly
well-characterized in the cerebellum-mediated adaptation of eye movement
reflexes. For example, in the vestibulo-ocular reflex (VOR), rapid
corrective eye movements are generated to offset movements of the
head, functioning like a motion-correcting camera. This reflex requires
tuning because, for example, the introduction of eyeglasses can alter
the relation between eye movement and resulting image motion across
the retina. Connecting to the present work, one can map the input
$r_{\text{in}}$ \Red{of Fig.~\ref{Fig1}B} to the head velocity, the output $r_{o}$ to the
eye velocity, $w^{*}$ to the desired VOR gain (i.e., the ratio
of eye to head velocity), and the teaching error to the "retinal slip" motion of the visual image on the retina \Red{\cite{bhasin2024synaptic}}. Learning
is then transferred from an initial site in the cerebellum (weight
$w_{1}$) to a late site of final storage in the vestibular nucleus
(weight $w_{2}$). Interestingly, to properly model the biological
circuit, one should make the climbing-fiber-driven error signals come
through discrete spikes rather than the smooth firing rate assumed
here. This provides an effective form of perturbation $\xi(t)$ that
can decrease the stability of the system if not compensated for by
decreasing the learning rate at the late site. Finally, we note that
a similar consolidation of learning has been shown to occur in the
striato-neocortical reinforcement learning system of the brain \cite{tachibana2022performance},
suggesting similar fundamental constraints on the speed of learning
may be applicable more broadly.

\Red{We have shown that the theory also can be applied to tuning the dynamics of a simple network model of the oculomotor neural integrator.
Temporal integration is a fundamental operation in the nervous system, where it is thought to mediate not only motor control, as focused upon here, but also the accumulation of evidence for decision making \cite{shadlen2013decision}. Within the eye movement system, loss of function of the oculomotor neural integrator is associated with eye movement disorders of gaze-holding, many of which may be due to improper tuning of the neural integrator by the cerebellum \cite{leigh2015neurology}.}

\Red{Our work differs from previous research on modeling the tuning of the oculomotor neural integrator in several ways. Early models used direct retinal slip input to the neural integrator to train it \cite{arnold1992neural, arnold1997oculomotor, xie1999spike}. A later model used corrective saccades to provide negative derivative information as a teaching signal \cite{macneil2011fine}, while another used a derivative signal generated in an unsupervised manner to tune the integrator \cite{federerZylberberg2018self}.  All of these models were single-stage learning models, in contrast to the present work. These differences highlight the novelty that our model brings, which integrates an adaptive cerebellar controller and a two-stage learning process to tune the neural integrator.}

\Red{To illustrate the Lyapunov framework, we focused on tuning the dynamics of the leading eigenvector of a linear integrator network. More generally, within linear network theory, additional non-integrating modes of the network could be tuned by including additional Purkinje cells. Further, we expect the core principles---fast learning at the cerebellum and slower consolidation in the cerebellar nuclei, as demonstrated in our model---should carry over to more complex, nonlinear networks. For such nonlinear networks, we speculate that the nonlinear tuning could be accomplished through having a set of Purkinje cells that serve as learned, nonlinear basis functions, with learning occurring due to changes in the weights of the many granule cell inputs to the Purkinje cells that, for simplicity, were not explicitly modeled here. 
Recent advances in adaptive control theory  \cite{boffi2021implicit} provide new techniques that may help with analyzing the tuning of nonlinear target circuits, not only in linearly parameterized nonlinear circuits where adaptive parameters affect the network state linearly but also in nonlinearly parameterized networks.}


Although systems consolidation \Red{is often} considered \Red{to be}
a \Red{very} slow process in several reported neural systems, this is not always
the case. For declarative memories, the late stage can consolidate
quickly if the new memories have features that are consistent with
the existing structure of knowledge in the late stage \cite{tse2007schemas,kumaran2016CLS}.
Recent evidence from songbird motor consolidation suggests that the
consolidation process may happen faster than originally thought, occurring
online in the daytime, and not necessarily requiring offline nighttime
processes \cite{tachibana2022performance}. Enforcing the stability
and convergence of consolidation in these scenarios may reveal constraints
on the speed of learning and consolidation, similar to what we have
found in the current work.

Besides the important implications for neuroscience experiments, the
framework we have provided here may have engineering applications.
Classically in adaptive control theory, the tracking error or prediction
error is directly used to tune the parameters of single-stage adaptive
controllers \cite{slotine1991}. Our work gives the insight that using
a two-stage adaptive controller can give flexibility in terms of having
robust storage \Red{of} memory at the final site, as well as an extra knob
to tune the speed of learning in a stable manner. In systems with delayed negative feedback that are
subject to inappropriate oscillations, such two-stage learning could
be used to avoid deleterious resonance effects. Recent
machine learning approaches have focused on using machine learning
to generate and control complex dynamical systems \cite{chen2018neuralode,hasani2022closed}. 
Given that artificial neural networks with online, adaptive learning are increasingly in
demand throughout society, including in safety-critical tasks, this suggests a compelling need to develop frameworks that guarantee
the stability of such algorithms. We hope that the principles of systems
consolidation and Lyapunov theory introduced here could help current
progress in this area by highlighting the need for real-time, continuously
adaptive systems that are safe and stable.


\appendix

\section{METHODS}

\subsection{Toy model simulation} 
For the simulation of the toy model, we simulated Eqs. (\ref{rO}-\ref{w2lr})
using the MATLAB solver ode45, which is a fifth-order Runge-Kutta
method. The learning rate of the first stage was set to a fixed value
$\eta_{1}=0.01$ \Red{$\text{s}^{-1}$} in all simulations shown in Fig.~\ref{Fig1}. $\eta_{2}$
was zero in the single-stage \Red{models (in Fig.~\ref{Fig1}C,E)}, $0.0003$ \Red{$\text{s}^{-1}$} in the two-stage model
with slow consolidation (\Red{in} Fig.~\ref{Fig1}\Red{D,F}, left), and $1$~\Red{$\text{s}^{-1}$} in the two-stage
\Red{models} with fast consolidation (in Fig.~\ref{Fig1}D,F, right). $r_{\text{in}}(t)$
was generated by a step function with amplitude $0.1$, smoothed \Red{using} the filter $100/(100s+1)$, \Red{where} $s$ \Red{represents} the Laplace \Red{transform} variable.
In Fig.~\ref{Fig1}\Red{C,D,} a sinusoidal perturbation with \Red{a very small} time-varying
amplitude \Red{$a(t)\sin(\omega_\text{a} t)$} where $\omega_\text{a}=0.1~\frac{\text{rad}}{s}$
was \Red{applied}. $a(t)$ was sampled at random from a
uniform distribution in the range \Red{{[}0,~0.004{]}} with a \Red{100} s sampling
period -- \Red{the time variation of $a(t)$} was not necessary for our core results, but was included
to illustrate the effect of a slow non-stationary amplitude $a(t)$.  \Red{In Fig.~\ref{Fig1}E,F, a signal-dependent perturbation $\xi(t)=0.7\sin(\omega_\text{sd} t) e(t)$ where $\omega_\text{sd}=0.2~\frac{\text{rad}}{s}$ was applied for all three panels. Parameters for the simulations shown in Fig.~\ref{Fig3} are provided in the main text and figure captions. }

\Red{\subsection{Simulation of the oculomotor neural integrator}
We simulated the dimensionless neural integrator Eqs.~(\ref{eqNI1}-\ref{eqNI4}) using the forward Euler method in MATLAB with 30 neurons,  and with $dt=0.05$, for $6800$ time units, then plotted the results 
by converting to dimensionalized units by setting
$\tau=1$~s~\cite{fisher2013modeling}.  The entries of $\mathbf{d}$, $\mathbf{k}_\text{PC}$, and $\mathbf{k}_\text{sacc}$ were chosen from a uniform distribution from $[0,1]$, and then normalized such that $\mathbf{d}^\top \mathbf{d}=1$, $\mathbf{d}^\top \mathbf{k}_\text{PC}=1$, and $\mathbf{d}^\top \mathbf{k}_\text{sacc}=1$. 
The weight matrix of the integrator was initialized to $ \mathbf{\Omega}= 1.5 \mathbf{k}_\text{sacc} \mathbf{d}^\top$ so that it has an unstable leading eigenvalue of $1.5$, and then small random Gaussian noise with standard deviation $0.001$ was added to the off-diagonal elements such that the other modes remain leaky but with some variation of decay timescales. $\mathbf{w}_\text{PC}$ was initialized as a vector of zeros.
 The learning rates for Fig.~\ref{Fig4}D were set to $\eta_1 = 0.014$, $\eta_2 = 1.5217\times10^{-4}$, and $\mu=0.5$. The signal-dependent sinusoidal perturbation $\xi(t)=\mu e(t) \sin(\omega_\xi t)$, with $\omega_\xi=0.0029$, was chosen to illustrate the potential instability in the system.
  The saccadic commands were generated as random-amplitude pulses, 
  with amplitudes sampled from a Gaussian distribution with a standard deviation of $48$. Pulses were generated from a Poisson process with a mean of 1 pulse per time unit.
   On top of this Poisson rate, whenever the eye position went out of the range $[-2.625, 2.625]$,
  an additional corrective saccade with an absolute amplitude of $61.2$ in the opposite direction was generated to bring the eye position back within this range, and the Poisson process for generating regular saccades was reset.
   In Fig.~\ref{Fig4}E, the learning rates $\eta_1$ and $\eta_2$ were interchanged to illustrate how a network with slow early site and fast late site can potentially become unstable.}
   
    \Red{The code to generate the results can be found at}
    
   \Red{ \url{https://github.com/alemi/LyapunovSystemConsolidation}.}

\section{STABILITY DEFINITIONS AND THEOREMS}

For ease of notation, we present the following definitions, lemmas,
and theorems from \cite{slotine1991} in the context of a general non-autonomous dynamical
system ${\dot{\bfx}=\frac{d{\bfx}}{dt}=f(\bfx,t)}$ for the state
vector $\bfx\in\mathbb{R}^{N}$ with equilibrium point ${\bfx_{eq}=\mathbf{0}}$.

\subsection{Definitions}

The formal definition of the most basic notion of stability in the
Lyapunov sense for a non-autonomous system is \begin{definition}
The equilibrium point $\mathbf{0}$ is \underline{stable} at $t_{0}$
if for any $R\geq0$, there exists a positive scalar $r(R,t_{0})$
such that 
\begin{itemize}
\item[] $\|x(t_{0})\|<r$ \quad{}$\Rightarrow$ \quad{}$\|x(t)\|<R$ \qquad{}$\forall t\ge t_{0}$. 
\end{itemize}
Otherwise, the equilibrium point $\mathbf{0}$ is \underline{unstable}.
If the scalar r in the above can be chosen independently of $t_{0}$,
i.e., if r = r(R), then the equilibrium point $\mathbf{0}$ is \underline{uniformly unstable}.
\end{definition} If the above conditions are true for the whole state
space, then the stability is \emph{global}; otherwise, the stability
is \emph{local}.

A more refined and desirable concept is asymptotic stability, which
not only ensures that the state stays in a ball of arbitrarily small
radius around the equilibrium but also provides a statement about
convergence to the equilibrium: \begin{definition} The equilibrium
point $\mathbf{0}$ is \underline{asymptotically stable} at $t_{0}$
if 
\begin{itemize}
\item it is stable 
\item $\exists r(t_{0})>0$ \quad{}such that \quad{}$\|x(t_{0})\|<r(t_{0})$
\quad{}$\Rightarrow$ \quad{}$\|x(t_{0})\|\rightarrow0$ as $t\rightarrow\infty$. 
\end{itemize}
In addition, if there exists a ball of attraction $\mathbf{B}_{\mathbf{R_{0}}}$,
whose radius is independent of $t_{0}$, such that any system trajectory
with initial states in $\mathbf{B}_{\mathbf{R_{0}}}$ converges to
$\mathbf{0}$ uniformly in $t_{0}$, then the equilibrium point $\mathbf{0}$
is \underline{uniformly asymptotically stable}. \end{definition}

\begin{definition} A scalar continuous function $L(\bfx)$ is said
to be \underline{locally positive definite} if $L(\mathbf{0})=0$
and, in a ball around the origin 
\begin{itemize}
\item[] $\bfx\neq\mathbf{0}$ \qquad{}$\Rightarrow$ \qquad{}$L(\bfx)>0$. 
\end{itemize}
If the inequality in the above is replaced with $L(\bfx)\geq0$, then $L(\bfx)$ is (locally) \underline{positive semi-definite}. If
$L(\mathbf{0})=0$ and the above property holds over the whole state
space, then $L(\bfx)$ is said to be \underline{globally positive definite}.
If $L(\bfx)$ is positive (semi-)definite, then $-L(\bfx)$ is \underline{negative (semi-)definite}.

The time-varying function $L(\bfx,t)$ is said to be positive definite
if $L(\mathbf{0},t)=0$ and there is a time-invariant positive definite
function $L_{0}(\bfx)$ such that 
\begin{itemize}
\item[] $\forall t\geq t_{0},$ \qquad{}$L(\bfx,t)\geq L_{0}(\bfx)$. 
\end{itemize}
\end{definition}

\vspace{1pt}
 \begin{definition} A scalar function $L(\bfx,t)$ is said to be
\underline{decrescent} if $L(\mathbf{0},t)=0$, and if there exists
a time-invariant positive definite function $L_{l}(\bfx)$ such that 
\begin{itemize}
\item[] $\forall t\geq0,\>$ \qquad{}$L(\bfx,t)\leq L_{l}(\bfx)$. 
\end{itemize}
\end{definition}

\vspace{1pt}
 \begin{definition} A function g is said to be \underline{uniformly continuous}
on $\left[0,\infty\right)$ if 
\begin{itemize}
\item[] $\forall R>0,\exists\eta(R),\forall t_{1}\geq0,\forall t\geq0,\quad such~that\qquad|t-t_{1}|<\eta\Rightarrow|g(t)-g(t_{1})|<R$. 
\end{itemize}
\end{definition} This uniformity in time means that one can always
find an $\eta$ which does not depend on the point $t_{1}$.

\subsection{Lemmas and theorems}

Below, we provide the theorems and lemmas needed to prove the stability
of the systems in the main text. For their proofs, consult \cite{slotine1991}.
\begin{theorem}[Lyapunov theorem for non-autonomous systems]

~~~~\textbf{{Stability}}\emph{:} If, in a ball $\mathbf{B}_{\mathbf{R_{0}}}$
around the equilibrium point $\mathbf{0}$, there exists a scalar
function $L(\bfx,t)$ with continuous partial derivatives such that 
\begin{enumerate}
\item $L$ is positive definite 
\item $\dot{L}=\frac{dL}{dt}$ is negative semi-definite 
\end{enumerate}
then the equilibrium point $\mathbf{0}$ is stable in the sense of
Lyapunov.\\

\textbf{{Uniform stability and uniform asymptotic stability}}\emph{:}
If, furthermore, 
\begin{enumerate}
\item[3.] $L$ is decrescent, 
\end{enumerate}
then the origin is uniformly stable. If condition 2 is strengthened
by requiring that $L$ be negative definite, then the equilibrium
point is uniformly asymptotically stable.\\

 \textbf{{Global uniform asymptotic stability}}\emph{:} If the ball
$\mathbf{B}_{\mathbf{R_{0}}}$ is replaced by the whole state space,
and condition 1, the strengthened condition 2, condition 3, and the
condition 
\begin{enumerate}
\item[4.] $L(\bfx,0)$ is radially unbounded, i.e., $L(\bfx,0)\rightarrow\infty$
as $\|\bfx\|\rightarrow\infty$ 
\end{enumerate}
are all satisfied, then the equilibrium point at $\mathbf{0}$ is
globally uniformly asymptotically stable. \label{theorem:Lyapunov}
\end{theorem}

\vspace{5pt}
 To prove asymptotic stability in cases where it is not easy to prove
negative definiteness of $\dot{L}$, we use the following Lyapunov-like
lemma, which is a variant of Barbalat's lemma \cite{slotine1991},
that requires the derivative of $L$ to have some additional smoothness
property to ensure $\dot{L}$ converges to zero: 
\begin{lemma} If a scalar function $L(x,t)$ satisfies the following
conditions 
\begin{itemize}
\item $L(x,t)$ is lower bounded 
\item $\dot{L}(x,t)$ is negative semi-definite 
\item $\dot{L}(x,t)$ is uniformly continuous in time 
\end{itemize}
then $\dot{L}(x,t)\to0$ as $t\to\infty$. \label{lemma:lyapunovlike}
\end{lemma} The first two conditions in the above Lemma imply that
$L$ has a finite limiting value $L_{\infty}$, but they do not guarantee
that $L$ will remain stationary at $L_{\infty}$ \cite{slotine1991}.
The addition of the third condition gives us the ability to conclude
that, in the limit $t\rightarrow\infty$, $L$ remains stationary
at $L_{\infty}$ and the convergence will be achieved.


\begin{acknowledgments}
We thank Jay Bhasin for helpful discussions. We acknowledge financial support from Simons Foundation Collaboration on the Global Brain grants 542989SPI and NC-GB-CULM-00002734 (MG, EA), NIH R01 NS104926 (MG, EA), and NIH R01 EY031972 (MG).
\end{acknowledgments}

\bibliography{ConsolidationBib}

\end{document}